# Title: Graphene for Terahertz Applications

**Authors:** Philippe Tassin[1], Thomas Koschny[1], Costas M. Soukoulis[1,2,*]

**Affiliations:**

[1]Ames Laboratory—U.S. DOE and Department of Physics and Astronomy, Iowa State University, Ames, IA 50011, USA.

[2]Institute of Electronic Structure and Lasers (IESL), FORTH, 71110 Heraklion, Crete, Greece.

*Correspondence to:  soukoulis@ameslab.gov.

**Main Text:**

Graphene is a one-atom-thick sheet of carbon atoms arranged in a honeycomb lattice. It was first obtained by exfoliation of graphite in 2004 [1] and has since evolved into a thriving research topic because of its attractive mechanical, thermal, and electrical properties [2-4]. Graphene's unique electrical properties derive from the relativistic nature of its quasiparticles, resulting in exceptionally high electron mobility [1]. Graphene promises to revolutionize many applications [2-4], ranging from solar cells and light-emitting devices to touch screens, photodetectors [4], microwave transistors [5], and ultrafast lasers [6].

Graphene has a number of special qualities that also make it desirable for devices manipulating terahertz waves:

*Atomic length scale:* Graphene sheets are extremely thin (< 1 nm) since they consist of a single continuous layer of $sp^2$-hybridized carbon atoms. This enables device miniaturization down to the atomic length scale.

*Tunability:* The electromagnetic properties of biased or photo-excited graphene are determined by the Drude-like electric response of the free carriers and the interband transitions between the lower and upper bands of the Dirac cone. Both are tunable through control of the Fermi energy, allowing for unprecedented tunability of electromagnetic structures made of this material.

*Large kinetic inductance:* The electric response of a graphene sheet is highly reactive, resulting in the existence of strongly localized, deeply subwavelength plasmons as well as the reduction of the resonance frequency of resonant structures made of patterned graphene rendering their size much smaller than the vacuum wavelength.

*Broadband gain:* Its linear band structure with Dirac point has implications for the population inversion in graphene after optical excitation: it is possible to create an inversion of the whole conical electronic band around the Dirac point up to the pump frequency, which should afford sizeable broadband gain practically all the way from terahertz to optical frequencies.

While there is an abundance of estimates from simplified theoretical models and experimental data for the DC properties of graphene from transport measurements, robust experimental data on the THz properties, especially the complex optical sheet conductivity, has increasingly become available only in the last few years. However, it is exactly this data that is essential to assess the potential performance of real graphene in its various envisioned applications [7]: The sheet resistivity, i.e., the real part of the inverse complex sheet conductivity, determines the "loss," which relates to the dissipative quality factor of graphene-based metamaterials as well as the propagation length of surface plasmons in graphene plasmonics; The sheet inductivity, i.e., the imaginary part of the inverse complex sheet conductivity divided by the angular frequency, determines the "confinement" or kinetic inductance, which translates to the saturation frequency—or maximum resonance frequency—of small-scale metamaterial resonators as well as to the "subwavelengthness" and lateral confinement of the surface plasmons.

A survey of *direct experimental measurements* of the THz sheet conductivity of graphene found in literature (blue diamonds) has been compiled in Fig. 1 (the detailed data is provided in Table S1). The best experimental graphene [8] at THz has a resistivity (dissipative loss) of 129 Ω and an inductivity (confinement) of 13 Ω/THz. Other recent THz experiments have reported 634 Ω and 21 Ω/THz [9] as well as 1428 Ω and 63 Ω/THz [10], which agree with earlier work [11]. Select experimental data from transport measurements (green triangles) is shown for comparison. Theoretical investigations of potential graphene applications have chosen different—mostly very optimistic—*assumptions* (red squares) for the sheet conductivity of graphene. Note that the theoretical loss estimates are generally about one order of magnitude below the experimental data. However, a number of theoretical studies obtain interesting device behavior if the sheet resistivity is roughly below 100 Ω. Examples include 139 Ω and 69 Ω/THz [12] as well as 84 Ω and 8.4 Ω/THz [13].

A lot of effort has been directed to producing more pristine graphene samples and, indeed, suspended graphene (minimizing disorder and phonons originating from the substrate), "cleaned" *in situ* by ohmic heating, has shown impressive mobilities exceeding $10^5$ cm²/(V·s), but at the expense of very limited carrier concentrations (bias). On the other hand, huge doping levels have been demonstrated by electrolytic gating, but this led in turn to much higher scattering.

Interestingly, the experimental data lines up within a rather narrow corridor (shaded blue) indicating that the variance in carrier concentration roughly trades higher confinement for higher loss in existing graphene samples. However, this should not be considered to indicate a particular dependence of the momentum relaxation time on carrier concentration, as the available samples are very different from one another. Nevertheless, it illustrates the current discrepancy between the experimentally realizable and the theoretically desired performance of graphene.

For metamaterial applications, graphene has to compete with metals. A 30-nm film of gold [7] has an experimentally measured sheet resistivity of the order of 1 Ω (yellow (Au) and grey (Ag) triangles in the Fig. 1), two orders of magnitude smaller than the theoretical lower limit of 30 Ω for free-standing graphene at room temperature [14]. The miniaturization advantage of graphene is marginal at terahertz frequencies, since the gold film is still much thinner than the free-space wavelength, and other constraints (e.g., minimum area to maintain a useful magnetic moment) require larger unit cells anyway. One major drawback of metals is their lack of control over carrier density. This is where graphene offers a substantial advantage, since its properties can be easily tuned by applying a gate voltage.

For plasmonics the main challenge is the short propagating length in graphene that is of the order of a few surface plasmon wavelengths at best in state-of-the-art experiments. However, tunability and confinement, in particular, may outweigh this limitation in the THz region as there are few alternatives (confinement of SPP on metals is very poor below optical frequencies).

Graphene is a fascinating material for THz applications with its strengths in atomic thickness, easy tunability and high kinetic inductance. The major challenge for these resonant high-frequency applications remains to overcome the dissipative loss, which might be less of a disadvantage for surface plasmonics on graphene than for graphene as a substitute for metals in metamaterials.

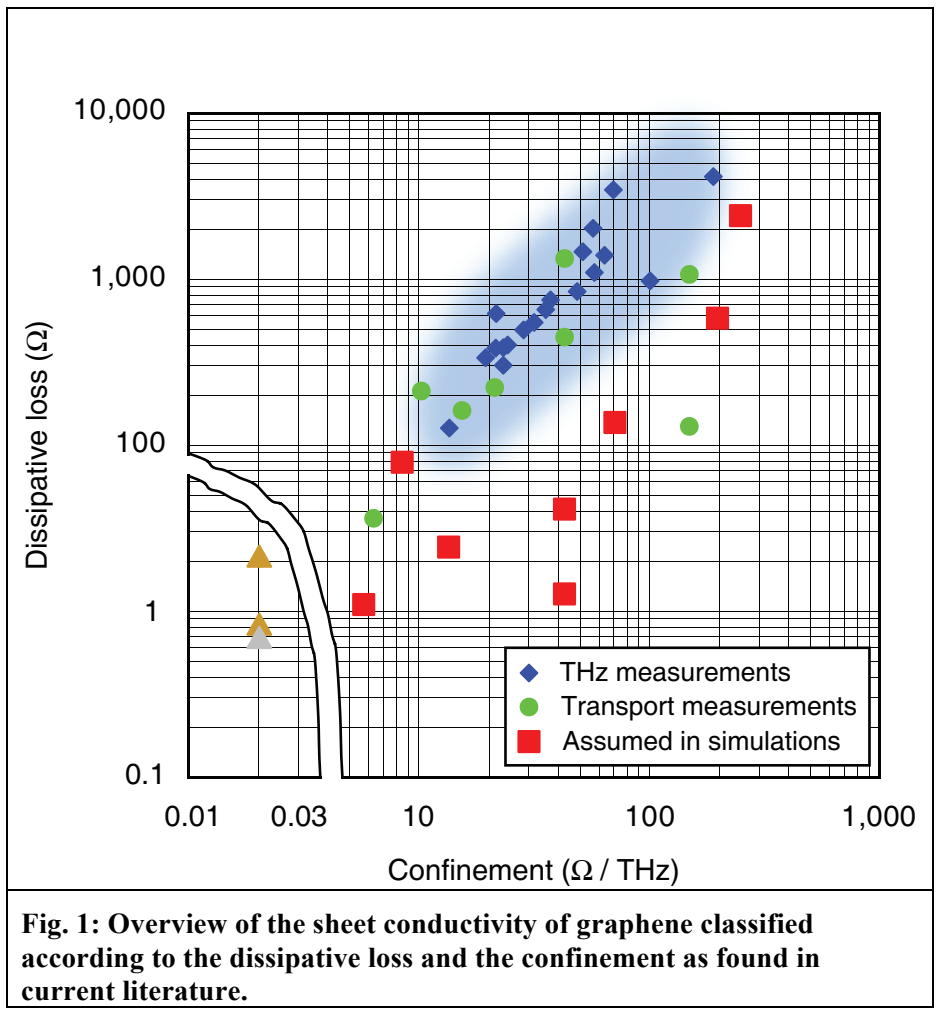

**Fig. 1: Overview of the sheet conductivity of graphene classified according to the dissipative loss and the confinement as found in current literature.**

15. Supported by Ames Laboratory, U. S. Department of Energy (Basic Energy Sciences, Division of Materials Sciences and Engineering) under contract DE-AC02-07CH1358; by ERC grant no. 320081 (PHOTOMETA); and by Office of Naval Research, Award No. N00014-10-1-0925.